\documentclass[%
    aps,
    prd,
    reprint,
    10pt,
    superscriptaddress,
    amsfonts,
    amssymb,
    amsmath,
    preprintnumbers,
    noshowpacs,
    tightenlines,
    floats,
    notitlepage,
    final,
    letterpaper,
    oneside,
    twocolumn,
    nofootinbib,
    longbibliography,
    ]{revtex4-1}
\pdfoutput=1
\usepackage[utf8]{inputenc}
\usepackage[T1]{fontenc}
\usepackage[british]{babel}
\usepackage[british,cleanlook,printdayon]{isodate}
\usepackage[Symbolsmallscale]{upgreek}
\usepackage{mathtools}
\usepackage{isomath}
\usepackage[protrusion,tracking,kerning,spacing,final,babel]{microtype}
\usepackage{physics}
\usepackage[final]{hyperref}
\hypersetup{%
    colorlinks=true,
    citecolor=red,
    linkcolor=red,
    urlcolor=red,
    }%
\usepackage{cleveref}
\usepackage{orcidlink}

\newcommand{\rme}{\mathrm{e}}
\newcommand{\rmi}{\mathrm{i}}

\begin{document}

\title{Analog vacuum decay from vacuum initial conditions}

\author{Alexander~C.~Jenkins\,\orcidlink{0000-0003-1785-5841}}
\altaffiliation{Corresponding author}
\email{alex.jenkins@ucl.ac.uk}
\affiliation{Department of Physics and Astronomy, University College London, London WC1E 6BT, United Kingdom}

\author{Jonathan~Braden\,\orcidlink{0000-0001-9338-2658}}
\affiliation{Canadian Institute for Theoretical Astrophysics, University of Toronto, 60 St.~George Street, Toronto, Ontario, M5S 3H8, Canada}

\author{Hiranya~V.~Peiris\,\orcidlink{0000-0002-2519-584X}}
\affiliation{Department of Physics and Astronomy, University College London, London WC1E 6BT, United Kingdom}
\affiliation{The Oskar Klein Centre for Cosmoparticle Physics, Department of Physics, Stockholm University, AlbaNova, Stockholm, SE-106 91, Sweden}

\author{Andrew~Pontzen\,\orcidlink{0000-0001-9546-3849}}
\affiliation{Department of Physics and Astronomy, University College London, London WC1E 6BT, United Kingdom}

\author{Matthew~C.~Johnson\,\orcidlink{0000-0003-4199-0314}}
\affiliation{Department of Physics and Astronomy, York University, Toronto, Ontario, M3J 1P3, Canada}
\affiliation{Perimeter Institute for Theoretical Physics, 31 Caroline St.~N, Waterloo, Ontario, N2L 2Y5, Canada}

\author{Silke~Weinfurtner\,\orcidlink{0000-0001-7642-5476}}
\affiliation{School of Mathematical Sciences, University of Nottingham, University Park, Nottingham NG7 2RD, United Kingdom}
\affiliation{Centre for the Mathematics and Theoretical Physics of Quantum Non-Equilibrium Systems, University of Nottingham, Nottingham NG7 2RD, United Kingdom}

\begin{abstract}
    Ultracold atomic gases can undergo phase transitions that mimic relativistic vacuum decay, allowing us to empirically test early-Universe physics in tabletop experiments.
    We investigate the physics of these analog systems, going beyond previous analyses of the classical equations of motion to study quantum fluctuations in the cold-atom false vacuum.
    We show that the fluctuation spectrum of this vacuum state agrees with the usual relativistic result in the regime where the classical analogy holds, providing further evidence for the suitability of these systems for studying vacuum decay.
    Using a suite of semiclassical lattice simulations, we simulate bubble nucleation from this analog vacuum state in a 1D homonuclear potassium-41 mixture, finding qualitative agreement with instanton predictions.
    We identify realistic parameters for this system that will allow us to study vacuum decay with current experimental capabilities, including a prescription for efficiently scanning over decay rates, and show that this setup will probe the \emph{quantum} (rather than thermal) decay regime at temperatures $T\lesssim10\,\mathrm{nK}$.
    Our results help lay the groundwork for using upcoming cold-atom experiments as a new probe of nonperturbative early-Universe physics.
\end{abstract}

\date{\today}
\maketitle

\section{Introduction}\label{sec:intro}

\begin{figure}[t!]
    \centering
    \includegraphics{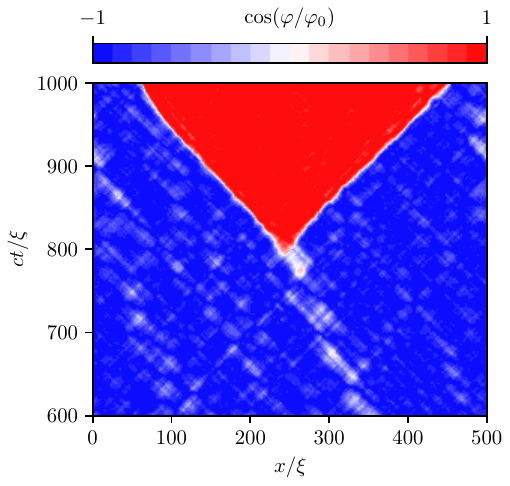}
    \caption{\label{fig:bubble-nucleation}
    Lattice simulation of vacuum decay in the 1D analog system.
    Nonlinear interactions between fluctuations around the false vacuum (blue) lead to the nucleation of a true vacuum bubble (red), which then expands relativistically.
    The simulation shown here corresponds to the blue curves in Fig.~\ref{fig:kg-charges}, and conserves the Hamiltonian of the effective relativistic theory to within $\sim10\%$ (see discussion in Sec.~\ref{sec:kg-charges}).}
\end{figure}

The decay of metastable `false vacuum' states via the nucleation of `true vacuum' bubbles (as illustrated in Fig.~\ref{fig:bubble-nucleation}) is a quintessential problem in nonperturbative quantum field theory~\cite{Coleman:1977py,Callan:1977pt,Coleman:1980aw,Linde:1981zj,Weinberg:2012pjx}.
This process has a broad range of applications in cosmology, including eternal inflation and multiverse scenarios~\cite{Guth:2007ng,Aguirre:2007gy,Aguirre:2007an,Feeney:2010jj,Feeney:2010dd}, electroweak baryogenesis~\cite{Kuzmin:1985mm,Cohen:1993nk,Morrissey:2012db}, Higgs vacuum decay~\cite{Ellis:2009tp,Degrassi:2012ry,Buttazzo:2013uya}, and the production of strong gravitational-wave signals~\cite{Kosowsky:1991ua,Kamionkowski:1993fg} (and potentially primordial black holes~\cite{Hawking:1982ga,Kodama:1982sf}) from bubble collisions.
These gravitational-wave signals in particular are a candidate source for the gravitational-wave background recently detected by various pulsar timing arrays, including NANOGrav~\cite{NANOGrav:2023gor,NANOGrav:2023hvm} and the European Pulsar Timing Array~\cite{EPTA:2023fyk,Antoniadis:2023xlr}, and are also one of the key obsevational targets of the planned space-based interferometer LISA~\cite{LISA:2017pwj,Caprini:2015zlo}.

Since the pioneering early work of Coleman and collaborators~\cite{Coleman:1977py,Callan:1977pt,Coleman:1980aw}, false vacuum decay (FVD) has primarily been studied using instanton methods, in which one obtains a semiclassical approximation of the decay rate by solving the equations of motion in imaginary time.
These methods are made tractable by imposing $O(d+1)$ symmetry on the resulting Euclidean `bounce' solutions which describe the bubble nucleation event (with $d$ the number of spatial dimensions).
However, this symmetry assumption is broken on dynamical and/or inhomogeneous spacetimes that are relevant to cosmology, and precludes us from studying interesting and observationally important issues such as correlations between multiple bubbles~\cite{Pirvu:2021roq,DeLuca:2021mlh}.
Furthermore, additional assumptions are required to interpret the instanton in real time; specifically, it is assumed that a critical bubble `appears' at some instant in time.
This prevents any study of  the precursors of such an event in terms of the real-time dynamics of the field.

Recently, a promising new method for addressing these questions has emerged: the use of ultracold atomic Bose gases as \emph{quantum simulators} of relativistic bubble nucleation~\cite{Opanchuk:2013lgn,Fialko:2014xba,Fialko:2016ggg,Braden:2017add,Billam:2018pvp,Braden:2019vsw,Billam:2020xna,Ng:2020pxk,Billam:2021qwt,Billam:2021nbc,Billam:2022ykl,Jenkins:2023npg}.
These systems exhibit coherent quantum behavior on scales that can be directly imaged in the laboratory, and can be manipulated into mimicking the dynamics of a Klein-Gordon field in a potential with true and false vacua.
Cold-atom experiments have already been successfully used to study discontinuous phase transitions in
quantum fields~\cite{Struck:2013ixy,Campbell:2016soc,Trenkwalder:2016qpt,Qiu:2020kzm,Song:2021pyy,Cominotti:2022jrj}, including nonrelativistic thermal vacuum decay~\cite{Zenesini:2023afv}.
Atomic simulators of \emph{relativistic} FVD are now under active development by several groups, offering the prospect of studying vacuum decay in real time and in a controlled and reproducible manner, with the promise of new insights that complement those from long-established Euclidean techniques.
These insights could have a transformative impact on our understanding of the early Universe, potentially helping to answer some of the most fundamental questions in cosmology, such as why there is more matter than antimatter~\cite{Kuzmin:1985mm,Cohen:1993nk,Morrissey:2012db}, and whether our observable Universe is embedded in a larger `multiverse'~\cite{Guth:2007ng,Aguirre:2007gy,Aguirre:2007an,Feeney:2010jj,Feeney:2010dd}.

Previous analyses of these analogs have focused on their \emph{classical} equations of motion, showing that these are equivalent to the Klein-Gordon equation for a relativistic field in the appropriate limit.
Here we go further by calculating the spectrum of \emph{quantum} vacuum fluctuations in the analog false vacuum state.
This fluctuation spectrum is a crucial input for lattice simulations of the cold-atom system, in which the fluctuations are represented as classical stochastic variables in order to obtain a semiclassical approximation of the decay process.
These simulations are our main theoretical tool for guiding the development of the analog experiments, and ultimately for helping us interpret the experimental data.

After describing our proposed analog system in Sec.~\ref{sec:bec}, we show in Sec.~\ref{sec:fluctuations} that the false-vacuum fluctuation spectrum matches that of a Klein-Gordon field on scales where the classical analogy holds.
This result was not guaranteed by the existing classical analogy, and thus provides further evidence for the suitability of this system as a relativistic analog.
After an exhaustive search of the cold-atom literature, we identify a homonuclear potassium-41 mixture as the most promising experimental setup, and in Sec.~\ref{sec:params} we present a realistic set of parameters for a 1D realization of this system.
This includes a protocol for scanning over parameters that allows us to vary the decay rate while keeping all other scales in the effective relativistic theory fixed.
In Sec.~\ref{sec:lattice} we then carry out a suite of semiclassical lattice simulations of this system, using our results for the fluctuation spectrum to generate realistic vacuum initial conditions.
We verify that the field undergoes exponential decay as expected, and that the decay rate scales exponentially with the amplitude of the initial fluctuations, in qualitative agreement with the instanton prediction.
Finally, in Sec.~\ref{sec:finite-temp} we explore the impact of finite temperatures on the decay rate, and argue that current experimental technologies can probe the regime of \emph{quantum} rather than thermal decays.
We summarize our results in Sec.~\ref{sec:summary}, and discuss avenues for further development of this work.

\section{The analog false vacuum}\label{sec:bec}

In this section we review the essential details of the analog FVD system we are interested in, as first proposed by \citet{Fialko:2014xba}, and subsequently studied in Refs.~\cite{Fialko:2016ggg,Braden:2017add,Billam:2018pvp,Braden:2019vsw,Billam:2020xna,Ng:2020pxk,Billam:2021qwt}.
This system consists of a two-component Bose-Einstein condensate (BEC), with each atomic species described by a complex bosonic field\footnote{Here and throughout, objects with hats denote quantum operators.}
    \begin{equation}
        \hat{\psi}_i(\vb*x)=\sqrt{\hat{n}_i(\vb*x)}\,{\exp}(\rmi\hat{\phi}_i(\vb*x)),\quad i=1,2.
    \end{equation}
The operators $\hat{\psi}_i^\dagger(\vb*x)$ and $\hat{\psi}_i(\vb*x)$ create and annihilate atoms of species $i$ in the position eigenstate $\ket{\vb*x}$, respectively.
Their amplitudes therefore determine the local number density of each species, $\hat{n}_i(\vb*x)=\hat{\psi}_i^\dagger\hat{\psi}_i$, while their phases $\hat{\phi}_i(\vb*x)$ encode coherent wavelike behavior and interference effects.
The dynamics of these fields are described by the Hamiltonian
    \begin{equation}
    \label{eq:hamiltonian}
        \hat{H}_0=\int_V\dd{\vb*x}\sum_i\hat{\psi}_i^\dagger\qty(-\frac{\hbar^2}{2m}\laplacian+\frac{1}{2}g\hat{\psi}_i^\dagger\hat{\psi}_i)\hat{\psi}_i,
    \end{equation}
    which consists of a nonrelativistic kinetic term for each species, as well as a quartic self-interaction of strength $g>0$ due to repulsive $s$-wave contact interactions between atoms.
This interaction sets the characteristic energy scale of the BEC, $E=gn$, where $n=\ev*{\hat{n}}$ is the mean number density.
The integral in Eq.~\eqref{eq:hamiltonian} is over a finite spatial volume $V$ that is either one- or two-dimensional, with the BEC confined tightly along the remaining dimensions, rendering them nondynamical.

We have specialized here to the case where both species have equal masses ($m_1=m_2=m$), equal intraspecies scattering ($g_{11}=g_{22}=g$), and zero interspecies scattering ($g_{12}=g_{21}=0$).
These conditions can be realized in practice by letting our two species be two different hyperfine states of the same atomic isotope, and applying an external magnetic field at the zero-crossing of a Feshbach resonance in the interspecies channel $g_{12}$~\cite{Fialko:2016ggg,Chin:2010fesh}.
Another possibility is to trap a single atomic species in a double-well potential; the atoms in each of the two wells then act as the two species, and only scatter with other atoms in the same well~\cite{Neuenhahn:2012dz,Tajik:2022lyt}.

The Hamiltonian~\eqref{eq:hamiltonian} excludes the usual external potential term that describes the trapping of the atoms along the extended direction(s).
Our proposed experiment uses a `box trap' which effectively approximates an infinite-well potential~\cite{Gaunt:2013box,Navon:2021mcf}, so that the given Hamiltonian is accurate inside the trap.
This is desirable for simulating relativistic physics as it maintains translation invariance in the interior region, with a near-homogeneous density profile.
The density rapidly tapers to zero at the walls of the trap on a characteristic scale called the \emph{healing length},\footnote{Note that this differs by a factor of $\sqrt{2}$ from the convention used by some authors.}
    \begin{equation}
        \xi=\frac{\hbar}{\sqrt{mgn}}.
    \end{equation}
For the experimental parameters we consider here, this scale is smaller than the size of the BEC by a factor of 500 (see Table~\ref{tab:parameters}).
We therefore treat the field as homogeneous with periodic boundaries throughout this paper, as in most previous studies of this system~\cite{Fialko:2014xba,Fialko:2016ggg,Braden:2017add,Braden:2019vsw,Billam:2020xna,Ng:2020pxk,Billam:2021qwt,Billam:2021nbc}.
(This setup is also a reasonable approximation to a 1D ring trap, as used in e.g.~Ref.~\cite{Banik:2021xjn}.)
Extending our results below to include the box trap and corresponding boundary conditions requires a calculation of the full spectrum of inhomogeneous eigenmodes, which has yet to be carried out for this system.
We will present this calculation and its impact on bubble nucleation in an upcoming companion paper.

The two condensed species are coupled via a linear interaction term in the Hamiltonian,
    \begin{align}
    \begin{split}
    \label{eq:hamiltonian-time-dependent}
        \hat{H}&=\hat{H}_0-\hbar\nu(t)\hat{H}_\mathrm{int},\quad\hat{H}_\mathrm{int}=\int_V\dd{\vb*x}(\hat{\psi}_1^\dagger\hat{\psi}_2+\hat{\psi}_2^\dagger\hat{\psi}_1),
    \end{split}
    \end{align}
    which allows atoms of species 1 to convert into species 2 (and vice-versa) at a rate $\nu$ that undergoes rapid modulation at some angular frequency $\omega$,
    \begin{equation}
        \hbar\nu(t)=\epsilon gn+\lambda\hbar\omega\sqrt{\epsilon/2}\cos(\omega t),
    \end{equation}
    where $\epsilon\ll1$ and $\lambda=\order{1}$ are dimensionless constants.
In the setup with two hyperfine states, this coupling is introduced by applying a modulated radio-frequency (rf) field; in the double-well case, $\nu$ instead represents the tunneling rate between the two wells.
We integrate out the fast oscillation to obtain an effective Hamiltonian $\hat{H}_\mathrm{eff}$ that is valid on timescales much longer than $\omega^{-1}$~\cite{Goldman:2014xja}.
At linear order in $\epsilon$, we find
    \begin{align}
    \begin{split}
    \label{eq:H_eff}
        \hat{H}_\mathrm{eff}&=\hat{H}_0+\epsilon gn\hat{H}_\mathrm{int}\\
        &+\frac{1}{4}\epsilon\lambda^2g\int_V\dd{\vb*x}\bigg(4\hat{\psi}_1^\dagger\hat{\psi}_2^\dagger\hat{\psi}_1\hat{\psi}_2-\hat{\psi}_1^\dagger\hat{\psi}_1^\dagger\hat{\psi}_1\hat{\psi}_1\\
        &\qquad-\hat{\psi}_2^\dagger\hat{\psi}_2^\dagger\hat{\psi}_2\hat{\psi}_2-\hat{\psi}_1^\dagger\hat{\psi}_1^\dagger\hat{\psi}_2\hat{\psi}_2-\hat{\psi}_2^\dagger\hat{\psi}_2^\dagger\hat{\psi}_1\hat{\psi}_1\bigg).
    \end{split}
    \end{align}
This time-averaged picture fails to capture the presence of Floquet instabilities induced in modes whose natural frequencies are close to the driving frequency $\omega$~\cite{Braden:2017add,Braden:2019vsw}.
One expects that setting $\omega$ sufficiently large (i.e., making the wavelengths of the unstable modes sufficiently short) will cause these instabilities to be quenched by damping effects on small scales; however, the exact nature of this process is still an open question.

\begin{figure}[t!]
    \centering
    \includegraphics{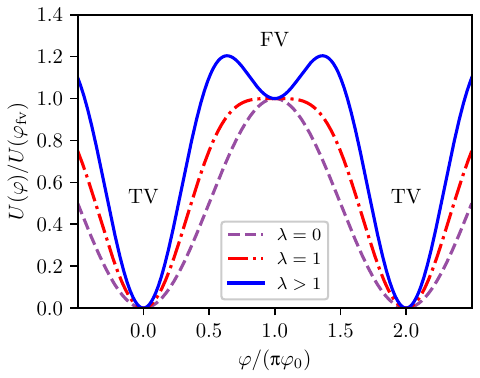}
    \caption{\label{fig:potential}
    Potential for the analog relativistic field $\hat{\varphi}$, as given by Eq.~\eqref{eq:potential}.
    There are stable `true vacuum' (TV) states at every even integer value of $\varphi/(\uppi\varphi_0)$.
    For $\lambda>1$ there are also metastable `false vacuum' (FV) states for every odd integer value.}
\end{figure}

\begin{figure*}[t!]
    \centering
    \includegraphics{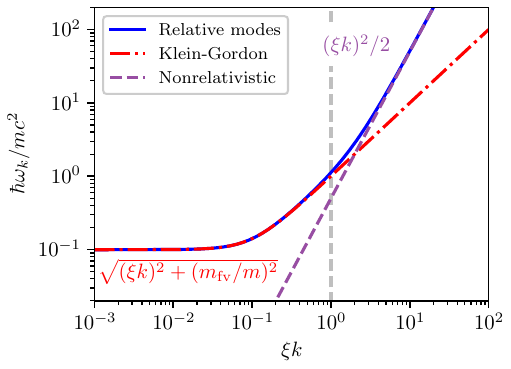}
    \includegraphics{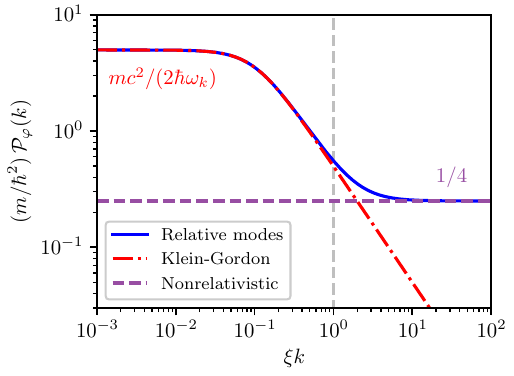}
    \caption{\label{fig:fluctuations}
    \emph{Left panel}: Dispersion relationship~\eqref{eq:dispersion-relation} for the relative modes of the analog system.
    \emph{Right panel}: Fluctuation power spectrum for the effective relativistic field $\hat{\varphi}$.
    Both quantities interpolate between being Klein-Gordon-like~\eqref{eq:kg-spectrum} in the IR ($\xi k\ll1$) and nonrelativistic~\eqref{eq:white-noise} in the UV ($\xi k\gg1$).
    The vertical dashed line in each panel indicates the crossover between these two regimes.
    Both use our fiducial parameters, given in Table~\ref{tab:parameters}.}
\end{figure*}

The relevance of the effective Hamiltonian~\eqref{eq:H_eff} for quantum simulation comes from considering the field
    \begin{equation}
    \label{eq:varphi-definition}
        \hat{\varphi}\equiv\varphi_0(\hat{\phi}_1-\hat{\phi}_2),\qquad\varphi_0\equiv\sqrt{\frac{\hbar^2n}{2m}},
    \end{equation}
    which is proportional to the relative phase between the two species.\footnote{%
    The normalization here is arbitrary at the classical level, but is chosen such that the quantum fluctuations in $\hat{\varphi}$ around the false vacuum exactly match those of the corresponding canonically-normalized Klein-Gordon field.}
On scales much larger than the healing length, the classical equation of motion for this degree of freedom is identical to that of a relativistic scalar field,
    \begin{equation}
    \label{eq:effective-kg}
        \qty(c^{-2}\partial_t^2-\laplacian)\varphi+U'(\varphi)=0,
    \end{equation}
    where we identify the `speed of light' as
    \begin{equation}
        c=\sqrt{gn/m}.
    \end{equation}
Note that in reality this is the sound speed of phonons in the BEC, which is roughly eleven orders of magnitude smaller than the speed of light in vacuum.
However, as we see below, it plays exactly the same role as the speed of light in the effective relativistic theory that emerges on large scales.

The potential appearing in Eq.~\eqref{eq:effective-kg} is
    \begin{equation}
    \label{eq:potential}
        U(\varphi)=4\epsilon\varphi_0^2\frac{m^2c^2}{\hbar^2}\qty[1-\cos(\varphi/\varphi_0)+\frac{1}{2}\lambda^2\sin^2(\varphi/\varphi_0)].
    \end{equation}
As shown in Fig.~\ref{fig:potential}, this contains a series of true vacua at $\varphi_\mathrm{tv}/\varphi_0=2j\uppi$, $j\in\mathbb{Z}$, and for $\lambda>1$, a series of false vacua at $\varphi_\mathrm{fv}/\varphi_0=(2j+1)\uppi$.
These correspond to the two atomic species being in phase and in antiphase, respectively; the linear coupling means that there is an additional energy density of order $\epsilon gn^2$ associated with being in antiphase, while the modulation generates an effective potential barrier that makes this state metastable.
Increasing the amplitude of the modulation via $\lambda$ creates a deeper potential barrier, and increases the mass of fluctuations in the false vacuum,
    \begin{equation}
    \label{eq:fv-mass}
        m_\mathrm{fv}^2=\frac{\hbar^2}{c^2}U''(\varphi_\mathrm{fv})=4\epsilon m^2(\lambda^2-1).
    \end{equation}

\section{Quantum fluctuations in the false vacuum}\label{sec:fluctuations}

We have reviewed the known result that, on scales much larger than the healing length, an atomic Bose-Bose mixture can reproduce the classical equation of motion of a Klein-Gordon field~\eqref{eq:effective-kg} with a false vacuum potential~\eqref{eq:potential}.
However, vacuum decay is inherently quantum-mechanical, so it is important to test whether these systems are also analogous at the quantum level.
Here we perform this test by calculating the power spectrum of fluctuations in the false vacuum state $\ket{\Omega_\mathrm{fv}}$,
    \begin{equation}
    \label{eq:power-spectrum}
        \mathcal{P}_\varphi(k)\equiv\ev{\hat{\varphi}_{\vb*k}^\dagger\hat{\varphi}_{\vb*k}^{\phantom{\dagger}}}{\Omega_\mathrm{fv}},
    \end{equation}
    where $\hat{\varphi}_{\vb*k}$ are the Fourier modes\footnote{%
    Note that we have \emph{discrete} Fourier frequencies, as we are working in a finite volume $V$.
    Our conventions for the Fourier transform and its inverse are $f_{\vb*k}=V^{-1/2}\int_V\dd{\vb*x}\rme^{-\rmi\vb*k\vdot\vb*x}f(\vb*x)$ and $f(\vb*x)=V^{-1/2}\sum_{\vb*k}\rme^{\rmi\vb*k\vdot\vb*x}f_{\vb*k}$.} of the effective relativistic field~\eqref{eq:varphi-definition}.
Below we find that, on scales much larger than the healing length ($\xi k\ll1$), this spectrum asymptotically matches that of the corresponding Klein-Gordon field,
    \begin{equation}
    \label{eq:kg-spectrum}
        \mathcal{P}_\varphi(k)\simeq\frac{\hbar c^2}{2\omega_k},\qquad\omega_k^2\simeq c^2k^2+\frac{c^4}{\hbar^2}m_\mathrm{fv}^2,
    \end{equation}
    with corrections suppressed by powers of ${(\xi k)}^2$ and $\epsilon$.

To derive this result, we adopt the standard mean-field approximation~\cite{Pethick:2008bec} in which each atomic field consists of small quantum fluctuations around a highly-occupied classical condensate wavefunction,
    \begin{equation}
        \hat{\psi}_1=\sqrt{n}\rme^{-\rmi\mu t/\hbar}+\updelta\hat{\psi}_1,\quad\hat{\psi}_2=-(\sqrt{n}\rme^{-\rmi\mu t/\hbar}+\updelta\hat{\psi}_2).
    \end{equation}
The factor $(-1)$ here reflects the fact that the two species are in antiphase in the false-vacuum state.
We expand around a homogeneous mean-field wavefunction, whose phase evolves at a rate set by the chemical potential, $\mu=(1+\epsilon)gn$.
To study the dynamics of the fluctuations, it is convenient to remove this time evolution with a canonical transformation $\hat{\psi}_i\to\rme^{\rmi\mu t/\hbar}\hat{\psi}_i$.
This modifies the Hamiltonian to
    \begin{equation}
        \hat{K}_\mathrm{eff}=\hat{H}_\mathrm{eff}-\sum_i\int_V\dd{\vb*x}\mu\hat{\psi}_i^\dagger\hat{\psi}_i.
    \end{equation}

Expanding this new Hamiltonian to quadratic order in the fluctuations, we find that it can be written as
    \begin{align}
    \begin{split}
    \label{eq:quadratic-hamiltonian}
        \hat{K}_\mathrm{eff}&=K_0+\hat{K}_++\hat{K}_-,\\
        \hat{K}_\pm&=\frac{1}{2}gn\sum_{\vb*k\ne\vb*0}\bigg\{[\xi^2k^2+2-(2\mp2)\epsilon]\hat{\psi}_{\vb*k}^{\pm\dagger}\hat{\psi}_{\vb*k}^\pm\\
        &\qquad\qquad+\qty[1-(1\mp1)\epsilon\lambda^2]\qty(\hat{\psi}_{\vb*k}^{\pm\dagger}\hat{\psi}_{-\vb*k}^{\pm\dagger}+\hat{\psi}_{\vb*k}^\pm\hat{\psi}_{-\vb*k}^\pm)\bigg\},
    \end{split}
    \end{align}
    with $K_0$ a constant energy offset associated with the mean-field solution, and separate terms $\hat{K}_\pm$ governing the total and relative fluctuation modes,\footnote{%
    Note that this is only true because we have truncated the Hamiltonian at quadratic order in the fluctuations.
    At higher order there are interactions between the total and relative modes, and these can in principle spoil the relativistic analogy if the fluctuations are sufficiently large.}
    \begin{equation}
        \hat{\psi}_{\vb*k}^\pm\equiv\frac{1}{\sqrt{V}}\int_V\dd{\vb*x}\rme^{-\rmi\vb*k\vdot\vb*x}\frac{1}{\sqrt{2}}(\updelta\hat{\psi}_1\pm\updelta\hat{\psi}_2),
    \end{equation}
    with the normalization chosen such that the modes obey canonical bosonic commutation relations.
The field we are interested in is defined solely in terms of the relative modes, and at linear order in the fluctuations is given by
    \begin{equation}
    \label{eq:varphi-relative-modes}
        \hat{\varphi}_{\vb*k}^{\phantom{\dagger}}=\frac{\rmi\hbar c}{2\sqrt{gn}}(\hat{\psi}_{\vb*k}^{-\dagger}-\hat{\psi}_{\vb*k}^-).
    \end{equation}
We can therefore ignore the dynamics of the total modes for now, given that they are decoupled in the linear regime.
(We return to them in Sec.~\ref{sec:finite-temp}, as they play a significant role in the presence of thermal noise.)

To calculate the power spectrum~\eqref{eq:power-spectrum}, we must determine the eigenstates of the relative Hamiltonian $\hat{K}_-$ and identify $\ket{\Omega_\mathrm{fv}}$ as the lowest-lying of these states.\footnote{%
    Restricting ourselves to linear fluctuations around the false vacuum means that the lower-lying states near the true vacuum are not in the spectrum.}
We can do this by writing the Hamiltonian in diagonalized form,
    \begin{equation}
    \label{eq:diagonalized-hamiltonian}
        \hat{K}_-=\sum_{\vb*k\ne\vb*0}\hbar\omega_k\hat{a}_{\vb*k}^\dagger\hat{a}_{\vb*k}^{\phantom{\dagger}},
    \end{equation}
    so that each normal mode, described by the ladder operators $\hat{a}_{\vb*k}^{\phantom{\dagger}}$, $\hat{a}_{\vb*k}^\dagger$, acts as an independent harmonic oscillator.
The false vacuum $\ket{\Omega_\mathrm{fv}}$ is then identified as the state annihilated by $\hat{a}_{\vb*k}$ for all wavenumbers $\vb*k$.
In Appendix~\ref{sec:bogoliubov} we identify the appropriate Bogoliubov transformation relating the normal modes to the relative atomic field modes $\hat{\psi}_{\vb*k}^-$, $\hat{\psi}_{\vb*k}^{-\dagger}$.
The energy associated with excitations of the normal modes is given by
    \begin{equation}
    \label{eq:dispersion-relation}
        \hbar\omega_k=\frac{1}{2}gn\sqrt{\xi^2k^2+4\epsilon(\lambda^2-1)}\sqrt{\xi^2k^2+4-4\epsilon(\lambda^2+1)},
    \end{equation}
    which, on scales much larger than the healing length ($\xi k\ll1$), reduces to the dispersion relation~\eqref{eq:kg-spectrum} of a Klein-Gordon field of the same false vacuum mass~\eqref{eq:fv-mass} we found in our classical analysis of the equations of motion.
We can directly evaluate the power spectrum~\eqref{eq:power-spectrum} by writing the Fourier modes $\hat{\varphi}_{\vb*k}$ in terms of the normal modes $\hat{a}_{\vb*k}$ and using standard ladder operator identities.
In the same IR limit as before we find Eq.~\eqref{eq:kg-spectrum}, which is exactly what we expect for the corresponding Klein-Gordon field.

We already know from our classical understanding of the system that the relativistic analogy breaks down on scales much smaller than the healing length ($\xi k\gg1$).
In this limit, we recover a white-noise fluctuation spectrum and the usual nonrelativistic dispersion relation,
    \begin{equation}
    \label{eq:white-noise}
        \mathcal{P}_\varphi(k)\simeq\frac{\hbar^2}{4m}=\mathrm{const.},\qquad\hbar\omega_k\simeq\frac{\hbar^2k^2}{2m}.
    \end{equation}
The former represents an excess of power at small scales compared to the Klein-Gordon spectrum~\eqref{eq:kg-spectrum}, due to nonrelativistic, high-momentum excitations of individual atoms.
The interpolation between this regime and the Klein-Gordon-like results on large scales is shown in Fig.~\ref{fig:fluctuations}.

\section{Experimental parameters}\label{sec:params}

Our results for the false vacuum power spectrum are a general feature of the modulated Bose-Bose mixture system described in Sec.~\ref{sec:bec}, regardless of any particular experimental realization.
In this section, we describe a concrete set of experimental parameters (summarized in Table~\ref{tab:parameters}) that is achievable with current cold-atom experiments, and which will allow us to probe the physics of relativistic vacuum decay.

As highlighted in Sec.~\ref{sec:bec}, among the key requirements for our system are that both atomic species have equal masses ($m_1=m_2$), equal intraspecies scattering lengths ($a_{11}=a_{22}$), and negligible interspecies scattering ($a_{12}=0$).\footnote{%
    These 3D scattering lengths $a_{ij}$ determine the corresponding 1D interaction strength, $g_{ij}=2\hbar\omega_\bot a_{ij}$, where $\omega_\bot$ is the frequency of the transverse harmonic potential, $V_\mathrm{trap}=\frac{1}{2}m\omega_\bot^2(y^2+z^2)$.}
It is easy to select equal masses by using two hyperfine states of the same atomic isotope (i.e., a homonuclear mixture).
However, the conditions on the scattering lengths are more difficult to arrange.
It is possible to set $a_{12}$ to zero by applying an external magnetic field at the zero-crossing of a Feshbach resonance~\cite{Chin:2010fesh}, but there is then no further freedom to tune $a_{11}$ and $a_{22}$ in order to set them equal to each other.
Fortunately, as pointed out by \citet{Fialko:2016ggg}, ${}^{41}\mathrm{K}$ (potassium-41) possesses a Feshbach resonance between the $\ket{F,m_F}=\ket{1,0}$ and $\ket{1,+1}$ states with a zero crossing at $B\simeq675.256\,\mathrm{G}$, where the condition $a_{11}\simeq a_{22}$ is realized naturally with a precision of $\sim1\%$.
We have performed an exhaustive search of other known Feshbach resonances in homonuclear mixtures of stable bosonic isotopes of the alkali metals (${}^7\mathrm{Li}$~\cite{Fialko:2016ggg,Hulet:2020li7}, ${}^{23}\mathrm{Na}$~\cite{Inouye:1998na23,Stenger:1999na23}, ${}^{39}\mathrm{K}$~\cite{Lysebo:2010fesh,Etrych:2023k39}, ${}^{85}\mathrm{Rb}$~\cite{Claussen:2003rb85,Blackley:2013rb85}, ${}^{87}\mathrm{Rb}$~\cite{Marte:2002rb87,Volz:2003rb87}, and ${}^{133}\mathrm{Cs}$~\cite{Chin:2004cs133,Lange:2009cs133}), and have not found any other interstate resonances where the condition $a_{11}\simeq a_{22}$ is satisfied at the zero-crossing of $a_{12}$.
The ${}^{41}\mathrm{K}$ resonance specified above is therefore the optimal candidate system for simulating relativistic vacuum decay.

\begin{table}[t!]
    \centering
    \begin{tabular}{l l}
        \hline\hline
        Parameter & Value \\
        \hline
        Atomic isotope & ${}^{41}\mathrm{K}$ (potassium-41) \\
        Atomic mass & $m=40.96\,\mathrm{u}=6.802\times10^{-26}\,\mathrm{kg}$ \\
        Hyperfine states & $\ket{F,m_F}=\ket{1,0},\ket{1,+1}$ \\
        Magnetic field & $B=675.256\,\mathrm{G}$ \\
        Scattering length (3D) & $a=60.24\,a_0=3.188\,\mathrm{nm}$ \\
        Healing length & $\xi=80\,a=0.2550\,\upmu\mathrm{m}$ \\
        Box trap length & $L=500\,\xi=127.5\,\upmu\mathrm{m}$ \\
        $\#$ atoms per species & $5000\le N\le25000$ \\
        Number density (1D) & $39.21\,\upmu\mathrm{m}^{-1}\le n\le196.1\,\upmu\mathrm{m}^{-1}$ \\
        Dimensionless density & $10\le\bar{n}\le50$ \\
        Transverse trap frequency & $3.04\,\mathrm{kHz}\le\omega_\bot/2\uppi\le15.2\,\mathrm{kHz}$ \\
        Scattering strength (1D) & $0.08\,\mathrm{peV}\,\upmu\mathrm{m}\le g\le0.4\,\mathrm{peV}\,\upmu\mathrm{m}$ \\
        Energy scale & $gn=15.69\,\mathrm{peV}$ \\
        Temperature scale & $gn/k_\mathrm{B}=182.1\,\mathrm{nK}$ \\
        Sound speed & $c=\sqrt{gn/m}=6.079\,\mathrm{mm}\,\mathrm{s}^{-1}$ \\
        Sound-crossing time & $L/c=20.98\,\mathrm{ms}$ \\
        Mean rf field & $\nu_0=59.59\,\mathrm{Hz}$ \\
        Inter-species coupling & $\epsilon=\hbar\nu_0/gn=2.5\times10^{-3}$ \\
        rf modulation amplitude & $\lambda=\sqrt{2}$ \\
        rf modulation frequency & $\omega\ge680\,c/\xi=2\uppi\times2.58\,\mathrm{MHz}$ \\
        False vacuum mass & $m_\mathrm{fv}=\sqrt{4\epsilon(\lambda^2-1)}\,m=0.1\,m$ \\
        \hline\hline
    \end{tabular}
    \caption{\label{tab:parameters}%
    List of fundamental and derived parameters for our proposed 1D cold-atom experiment.
    Here $\mathrm{u}=1.661\times10^{-27}\,\mathrm{kg}$ is the unified atomic mass unit and $a_0=5.292\times10^{-11}\,\mathrm{m}$ is the Bohr radius.
    The scattering length $a$ quoted here is the mean of the two intrastate scattering lengths; the difference is $\sim1\%$ (cf.~Fig.~\ref{fig:feshbach}).
    The number density $n$ and scattering strength $g$ are scanned over by varying the number of atoms of each species $N$ and the harmonic trap frequency $\omega_\bot$ respectively, while holding the energy scale $gn$ constant.}
\end{table}

\begin{figure}[t!]
    \centering
    \includegraphics{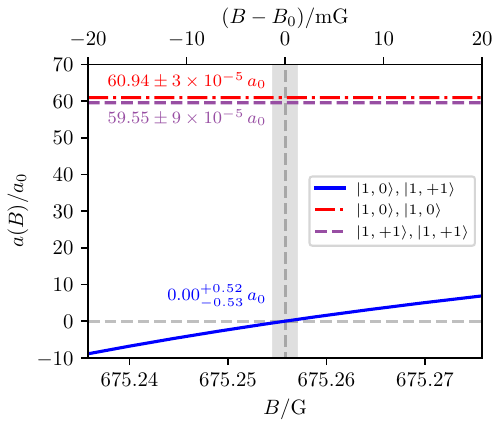}
    \caption{\label{fig:feshbach}
    The three scattering lengths $a_{11}$, $a_{22}$, $a_{12}$ of our proposed homonuclear ${}^{41}\mathrm{K}$ mixture as a function of magnetic field strength.
    The quoted values and gray shaded region correspond to $\pm2\,\mathrm{ppm}\approx\pm1.4\,\mathrm{mG}$ either side of the zero-crossing, as given by Ref.~\cite{Lysebo:2010fesh}.}
\end{figure}

The main technical challenge with this setup is that the resonance has a width of only $155.8\,\mathrm{mG}$~\cite{Lysebo:2010fesh}, necessitating a very high level of magnetic field stability in order to stay at the zero-crossing of $a_{12}$, as illustrated in Fig.~\ref{fig:feshbach}.
Nonetheless, this level of stability is achievable with current experimental technologies.
In particular, \citet{Borkowski:2023ppm} have recently demonstrated magnetic field stability at the level of $\sim2\,\mathrm{ppm}$ in a cold-atom experiment.
For our proposed system this corresponds to $\abs{a_{12}}\le0.53\,a_0$ (where $a_0=5.292\times10^{-11}\,\mathrm{m}$ is the Bohr radius).
This is less than $1\%$ of the mean intrastate scattering length $a=60.24\,a_0$, which should be sufficient precision for our purposes.

Given the 3D scattering properties of the two atomic species, the behavior of the effective 1D system is set by the number of condensed atoms, the size of the trap along the elongated and transverse directions, and the strength and modulation of the applied radio-frequency field.
We have explored this parameter space with the goal of maximizing the natural condensate energy scale $gn$ relative to the thermal energies $k_\mathrm{B}T$ that can be achieved in current experiments, as this will allow us to investigate the regime of quantum (rather than thermal) decays.
At the same time, we have ensured that this energy scale is not so high that transverse modes of energy $\hbar\omega_\bot$ are excited, where $\omega_\bot$ is the frequency of the harmonic trapping potential in the transverse directions.
(We plan to test this explicitly in future work with 3D simulations that resolve the transverse directions.)

In order to facilitate comparisons with instanton predictions (which are challenging to calibrate at any single point in parameter space), it is useful to vary the system parameters to scan over a broad range of bubble nucleation rates.
The instanton decay rate per unit volume in this model scales as
    \begin{equation}
    \label{eq:instanton-rate}
        \log(\Gamma/V)\propto-\epsilon^\frac{1-d}{2}\,\bar{n},
    \end{equation}
    where $\bar{n}\equiv\xi^d n$ is the dimensionless condensate number density (i.e., the number of atoms in a region of size equal to the healing length).
In $d=1$ dimensions the dependence on $\epsilon$ vanishes, and the decay rate is thus primarily controlled by $\bar{n}$.
This parameter also sets the size of fluctuations in the field relative to the characteristic value $\varphi_0$,
    \begin{equation}
        \sigma_\varphi^2/\varphi_0^2\propto1/\bar{n}.
    \end{equation}
We find that it is possible to vary $\bar{n}$ while keeping the energy scale $gn$ (and therefore all other dimensionless parameters of the system) fixed, by simultaneously increasing the number of atoms of each species $N$ and decreasing the transverse trapping frequency $\omega_\bot$.
This allows us to perform a controlled test of how the bubble nucleation rate scales with the amplitude of the initial fluctuations.

Our proposed parameters are summarized in Table~\ref{tab:parameters}.
We vary $\bar{n}$ by a factor of $5$, which is sufficient to see a significant variation in the decay rate.
As we show in Sec.~\ref{sec:finite-temp} below, the energy scale $gn$ here is large enough that the quantum-decay regime is readily accessible to current or near-future experiments.

\begin{figure}[t!]
    \centering
    \includegraphics{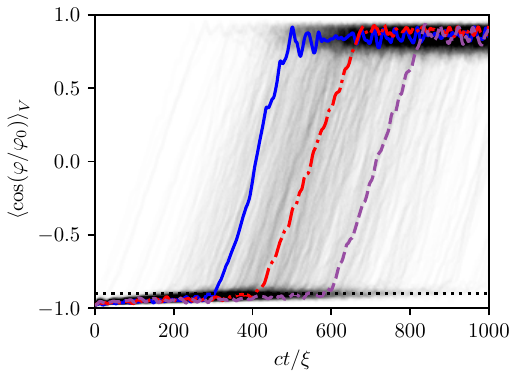}
    \caption{\label{fig:survival}
    Individual random realizations of vacuum decay in our $\bar{n}=35$ ensemble, showing how the volume-averaged cosine of the relative phase field evolves over time.
    Each curve corresponds to an independent simulation, which oscillates near $\ev{\cos(\varphi/\varphi_0)}_V=-1$ until a true vacuum bubble nucleates, at which point the trajectory grows until it saturates near $\ev{\cos(\varphi/\varphi_0)}_V=+1$.
    The colored curves are three randomly-selected trajectories, highlighted to illustrate the typical behavior.
    The black dotted line shows our empirically-determined decay threshold for this ensemble, as found using the procedure described in the main text.}
\end{figure}

\section{Lattice simulations}\label{sec:lattice}

Part of the value of our results on the vacuum power spectrum in Sec.~\ref{sec:fluctuations} is that they can be used as an input for semiclassical lattice simulations of the cold-atom system.
These simulations are a powerful tool for exploring the real-time dynamics of bubble nucleation, and are a crucial ingredient for developing and interpreting analog FVD experiments.
The key idea is to encode the nonclassical nature of the problem in the \emph{initial conditions} of the simulation, by drawing an ensemble of random field realizations that sample vacuum fluctuations around the homogeneous false vacuum state~\cite{Braden:2018tky}.
These realizations are then evolved forward by numerically integrating the classical equations of motion.
This approach is widely used in the context of atomic physics and quantum optics (where it is referred to as the `truncated Wigner approximation'~\cite{Werner:1997twa,Steel:1998twa,Blakie:2008vka}), and also underpins cosmological lattice simulations of inflation and preheating~\cite{Khlebnikov:1996mc,Khlebnikov:1996wr,Prokopec:1996rr,Khlebnikov:1996zt,Felder:2000hq,Rajantie:2000nj,Bond:2009xx,Amin:2010dc,Amin:2011hj,Clough:2016ymm,Figueroa:2020rrl,Figueroa:2021yhd} as well as vacuum decay~\cite{Braden:2018tky,Hertzberg:2020tqa,Batini:2023zpi}.

It is common for lattice simulations of cold-atom systems to initialize the fluctuations using a white-noise power spectrum~\eqref{eq:white-noise}~\cite{Fialko:2014xba,Fialko:2016ggg,Billam:2018pvp,Braden:2019vsw}, particularly in situations where the processes of interest are insensitive to the precise form of this spectrum.
Bubble nucleation, however, is extremely sensitive to the statistics of the initial fluctuations, as different initial states can decay at exponentially different rates.
(For example, we see from Eq.~\eqref{eq:instanton-rate} that there is an exponential sensitivity on $\bar{n}$.)
The vacuum fluctuation spectrum derived above is therefore a crucial ingredient for realistic simulations of analog vacuum decay.

In this section we use a suite of lattice simulations to study bubble nucleation from vacuum initial conditions in the 1D cold-atom system described in Sec.~\ref{sec:params}.
We extract decay rates for different values of the fluctuation-amplitude parameter $\bar{n}$, and verify that the rates depend exponentially on this parameter, in agreement with the scaling found in the instanton approach.
We perform the same test with white-noise initial conditions, and find decay rates that are globally larger than in the vacuum case.
This confirms that vacuum decay in semiclassical lattice simulations is indeed sensitive to the statistics of the initial fluctuations, and that for the cold-atom system these must be correctly specified using Bogoliubov theory, as we have done here.
We additionally investigate the conservation of the Noether charges of the effective Klein-Gordon theory in our simulations of the cold-atom system, as these are a useful diagnostic for the faithfulness of the relativistic analogy.

\begin{figure*}[t!]
    \centering
    \includegraphics{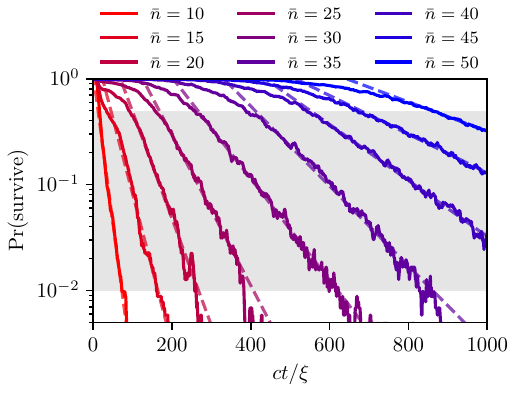}
    \includegraphics{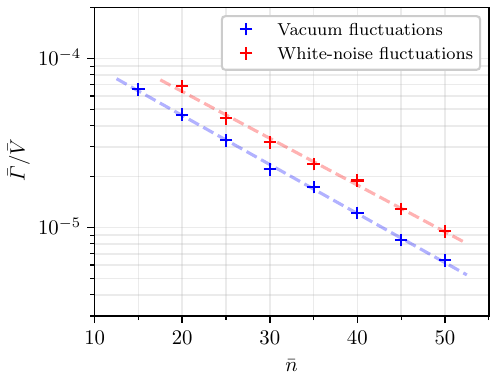}
    \caption{\label{fig:decay-rates}
    \emph{Left panel}: Survival probability for the false vacuum state as a function of time, as estimated using ensembles of $1024$ simulations for each curve.
    We scan over the dimensionless number density $\bar{n}=\xi n$ to probe a broad range of decay rates.
    The gray shaded region ($0.01\le\Pr(\mathrm{survive})\le0.5$) is used to fit an exponential decay rate $\Gamma$ for each curve (shown as dashed lines).
    \emph{Right panel}: Dimensionless decay rate per unit volume as a function of $\bar{n}$, computed for both vacuum and white-noise initial conditions.
    Both curves are well-described by a linear fit, as expected from instanton calculations.
    The white-noise case consistently gives faster decays despite the smaller initial phase fluctuations, due to this being an excited state of the system.}
\end{figure*}

\subsection{Code setup}

We use a Fourier pseudospectral code with an eighth-order symplectic time-stepping algorithm~\cite{Yoshida:1990zz} (see Appendix~\ref{sec:numerics} for details), and work in units where the atomic mass $m$, healing length $\xi$, and sound speed $c$ are set to unity (which is equivalent to also setting $\hbar=gn=1$).
Our simulations work at the level of the time-dependent Hamiltonian~\eqref{eq:hamiltonian-time-dependent}, resolving the modulation of the interspecies coupling so that we can test for the emergence of the effective time-averaged dynamics.

We simulate a system with the experimental parameters specified in Table~\ref{tab:parameters}.
In code units, this setup is realized by evolving a periodic region of volume $V/\xi^d=L/\xi=500$, and setting $\epsilon=2.5\times10^{-3}$ and $\lambda=\sqrt{2}$ so that the false vacuum mass is $m_\mathrm{fv}/m=0.1$.
We additionally set the dimensionless modulation frequency to $\omega\xi/c=680$, which is sufficiently large that the Floquet instability bands are above the Nyquist frequency for all of our simulations.
This allows us to model the expected experimental situation where these instabilities are damped by the small-scale dynamics of the BEC, and do not affect the evolution of the IR modes; the actual experimental value of $\omega$ is unimportant so long as the Floquet instabilities are quenched.
Our simulations use $2048$ lattice sites and a timestep that is $1/16$ times the modulation period $2\uppi/\omega$, giving spatial and temporal resolution of $\Updelta x/\xi\approx0.244$ and $c\Updelta t/\xi\approx5.77\times10^{-4}$, respectively.
In Appendix~\ref{sec:numerics} we show that our results are numerically converged at this resolution, and that the Noether charges of the cold-atom Hamiltonian~\eqref{eq:hamiltonian-time-dependent} are conserved to within a few parts per billion.

\subsection{Bubble nucleation rates}

We extract decay rates for the analog system using ensembles of $1024$ simulations, with each simulation corresponding approximately to a different possible classical history drawn from the path integral describing the full evolution of the many-body quantum state.
We initialize each simulation as the homogeneous false vacuum $\varphi=\uppi\varphi_0$ plus independent random draws of the vacuum fluctuations $\updelta\hat{\varphi}$.
We treat the latter as a zero-mean Gaussian random field with a power spectrum that (as shown in Fig.~\ref{fig:fluctuations}) interpolates between a relativistic spectrum in the IR and a white-noise spectrum in the UV.
We have checked that this power spectrum remains statistically stationary over time by averaging over the ensemble of nondecayed trajectories, effectively testing that our initial state is indeed an eigenstate of the Hamiltonian near the false vacuum.

As well as the relative phase, we also initialize the relative density and the total phase and density using random draws from their corresponding vacuum spectra.
It is crucial to initialize all four fields in this way to correctly capture the vacuum state.
For example, neglecting the relative density fluctuations corresponds to initializing the effective Klein-Gordon field with zero momentum everywhere, when in fact this momentum field should also contain vacuum fluctuations.
In practice, we initialize the total and relative atomic field modes in our code, which is equivalent at the linear level to working in terms of the density and phase fields.

We find that it is crucial that the positive- and negative-momentum Fourier modes $\psi_{\vb*k}$ and $\psi_{-\vb*k}$ are \emph{not} treated as statistically independent random variables.
Instead, one must draw the positive- and negative-momentum normal modes $a_{\vb*k}$, $a_{-\vb*k}$ independently, and then obtain the Fourier modes of the atomic fields via a reverse Bogoliubov transformation.
This induces a nontrivial correlation between $\psi_{\vb*k}$ and $\psi_{-\vb*k}$ that appropriately captures the quantum statistics of the false vacuum state.
Failing to include these correlations in the initial conditions puts the system into an excited state that nucleates bubbles much more rapidly than the false vacuum state, and much more even than the white-noise state.

\begin{figure*}[t!]
    \centering
    \includegraphics{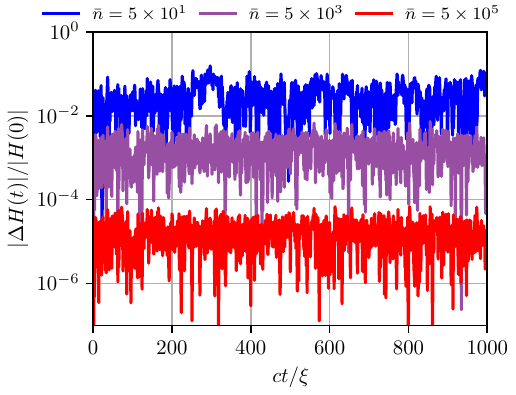}
    \includegraphics{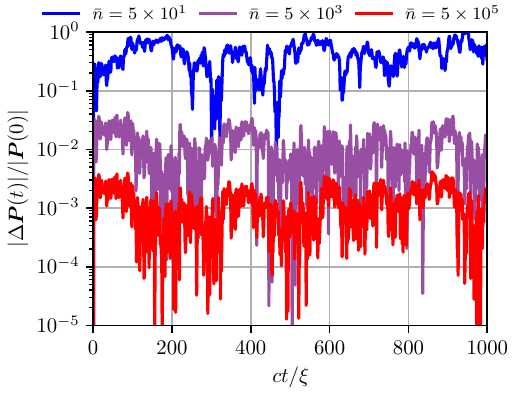}
    \caption{\label{fig:kg-charges}
    Fractional violation of the Klein-Gordon charges~\eqref{eq:kg-charges} as a function of the dimensionless BEC number density $\bar{n}$.
    The initial fluctuations in each simulation are identical except for an overall $\sim\bar{n}^{-1/2}$ scaling.
    The level of violation is roughly stationary throughout each simulation, and approaches zero for large $\bar{n}$, despite the nonrelativistic behavior of the system on small scales.
    }
\end{figure*}

We truncate all of the fluctuation spectra at a maximum wavenumber of $\xi k_\mathrm{UV}\approx3.22$, which is a factor of 4 smaller than the Nyquist frequency of our simulations, $\xi k_\mathrm{Nyq}=\uppi\xi/\Updelta x\approx12.9$.
Evidence from pure Klein-Gordon lattice simulations~\cite{Braden:2022odm} suggests that changing this cutoff modifies the decay rate in a way that can be absorbed into a renormalization of the bare model parameters.
We leave a detailed investigation of this effect in the analog system for future work, and here use a fixed UV cutoff for all of our simulations.

The amplitude of the fluctuations relative to the homogeneous value of the field is set by the dimensionless number density $\bar{n}$, which we scan over in the experimentally accessible range $10\le\bar{n}\le50$.
We measure a decay rate from each ensemble of simulations by counting the number of nondecayed trajectories as a function of time, dividing by the total number of simulations to obtain an estimate of the time-dependent survival probability.
In doing so, it is necessary to choose a definition for when an individual realization has decayed.
We do this by setting a threshold on the volume average of the cosine of the relative phase, $\ev*{\cos(\varphi/\varphi_0)}_V$.
This quantity fluctuates near to $-1$ in the false vacuum, and grows rapidly after a bubble nucleates before saturating near $+1$ once the transition has percolated, as illustrated in Fig.~\ref{fig:survival}.
We compute the decay threshold separately for each ensemble as the lowest possible value of $\ev*{\cos(\varphi/\varphi_0)}_V$ for which no more than 1\% of the simulations cross back below the threshold in any given timestep.\footnote{%
    A more obvious choice would be to allow zero downward crossings through the threshold, as this would capture the notion that vacuum decay is an irreversible process.
    However, we find that enforcing zero downward crossings makes the algorithm easily confused by small fluctuations in $\ev*{\cos(\varphi/\varphi_0)}_V$, and results in a choice for the threshold that is far too conservative.
    Manual inspection of the results with a 1\% allowance for downward crossings confirms that this accurately captures the common-sense notion of when the field has decayed (e.g. see Fig.~\ref{fig:survival}).
    We have checked that varying this allowed fraction between 0.5\% and 2\% does not significantly impact our measured decay rates.}

Our resulting estimates of the survival probability are shown in the left panel of Fig.~\ref{fig:decay-rates}.
As expected, the ensembles with smaller $\bar{n}$, and therefore larger initial fluctuations, decay on much shorter timescales.
After an initial transient, each ensemble reaches a regime of exponential decay,
    \begin{equation}
        \Pr(\mathrm{survive})\sim\exp(-\Gamma t).
    \end{equation}
We fit a decay rate $\Gamma$ to each curve, restricting the fit to survival probabilities between 50\% and 1\% in order to exclude the nonexponential regime at early times and noisy small-number statistics at late times, respectively.
The resulting decay rates (in dimensionless units, and measured per unit volume) are shown in blue in the right panel of Fig.~\ref{fig:decay-rates}, and are well-described by an exponential scaling with respect to $\bar{n}$, in qualitative agreement with the instanton prediction~\eqref{eq:instanton-rate}.

It is important to note however that the proportionality constant linking $\log(\Gamma/V)$ and $\bar{n}$ does \emph{not} agree with the instanton prediction; our simulations decay significantly faster than predicted in the instanton approach.
This same behavior has been observed in pure Klein-Gordon lattice simulations~\cite{Braden:2018tky}, and is an expected consequence of performing instanton calculations using the bare lattice parameters, rather than the renormalized theory~\cite{Braden:2022odm}.
It is also worth pointing out that our instanton calculations are based on the effective Klein-Gordon theory, rather than the full analog system, and therefore neglects effects such as the excess small-scale power identified in Sec.~\ref{sec:fluctuations}.
We plan to explore these issues in the context of the analog system in future work.

As well as our simulations using vacuum initial conditions, we carry out a suite of simulations using white-noise initial conditions.
This corresponds to the nonrelativistic UV limit~\eqref{eq:white-noise} of the full power spectrum derived from Bogoliubov theory, and matches the prescription used by several previous studies of vacuum decay in cold-atom analog systems~\cite{Fialko:2014xba,Fialko:2016ggg,Billam:2018pvp,Braden:2019vsw}.
The resulting decay rates are shown in red in the right panel of Fig.~\ref{fig:decay-rates}.
These are fit only to survival probabilities between $20\%$ and $1\%$, as we find that it takes longer for these initial states to settle into a period of steady exponential decay.
We see that, while the resulting decay rates also follow the expected exponential scaling with $\bar{n}$, they are globally larger for white-noise initial conditions than for the vacuum case, despite the fact that the actual amplitudes of the fluctuations are smaller in the IR in the white-noise case (compare the blue and purple curves in Fig.~\ref{fig:fluctuations}).
We interpret this as evidence that white-noise fluctuations correspond to an excited state of the analog system, and thus lead to faster decays, on average, than the vacuum initial conditions we have derived here.

Note that this does not imply that the white-noise spectrum is somehow unphysical.
In fact, such a spectrum is the vacuum state for an alternative system with zero atomic scattering, $g=0$.
The enhanced decay rates shown in red in Fig.~\ref{fig:decay-rates} can thus be interpreted as being due to a mismatch between the Hamiltonian describing the initial conditions and the Hamiltonian describing the time evolution.

\subsection{Verifying Klein-Gordon behavior}\label{sec:kg-charges}

While our results for the decay rates are in broad agreement with our expectations for relativistic vacuum decay, we can also directly test whether the relative phase field $\varphi$ is indeed analogous to a relativistic Klein-Gordon field by computing the Noether charges for the corresponding Klein-Gordon theory,
    \begin{align}
    \begin{split}
    \label{eq:kg-charges}
        H&=\int_V\dd{\vb*x}\qty[\frac{1}{2c^2}\dot{\varphi}^2+\frac{1}{2}\abs{\grad\varphi}^2+U(\varphi)],\\
        \vb*P&=-\int_V\dd{\vb*x}\frac{1}{c}\dot{\varphi}\grad\varphi.
    \end{split}
    \end{align}
Since the Noether charges for the underlying nonrelativistic Hamiltonian are conserved with extremely high precision in our simulations (see Appendix~\ref{sec:numerics}), any nonconservation of the Klein-Gordon charges~\eqref{eq:kg-charges} should be interpreted as being due to limitations of the relativistic analogy, rather than numerical errors.

In Fig.~\ref{fig:kg-charges} we show the violation of these charges for a series of simulations with a broad range of dimensionless number densities $\bar{n}$.
We find that violations in the Klein-Gordon energy and momentum are roughly stationary over time, and reach a regime where they scale like $\abs{\upDelta H}/\abs{H}\sim\bar{n}^{-1}$ and $\abs{\upDelta\vb*P}/\abs{\vb*P}\sim\bar{n}^{-1/2}$ respectively, so that in the limit of small fluctuations the analogy holds with high accuracy.
However, in the experimentally accessible regime $\bar{n}\in[10,50]$ that we are interested in here, the violation is on the order of at least a few percent in the energy.
In the momentum, the relative errors reach order unity, although this reflects the fact that the total momentum of the field averaged over the entire volume $V$ is intrinsically close to zero.

While we do not believe these errors invalidate the mapping onto the Klein-Gordon theory, further improvements in the accuracy of the analog may be possible.
Specifically, so far we have ignored the backreaction of the fluctuations onto the mean-field dynamics, which would modify this mapping in a way that could plausibly be absorbed into a renormalization of the parameters of the effective Klein-Gordon theory.
(Similar effects have recently been investigated in the case of pure Klein-Gordon theory~\cite{Braden:2022odm}.)
This would be consistent with our finding that the level of charge violation scales with the fluctuation amplitudes.
We conjecture that accounting for these corrections and identifying the appropriate Klein-Gordon parameters could substantially improve the level of charge violation over that shown in Fig.~\ref{fig:kg-charges}, and also bring our decay rates into closer quantitative agreement with the instanton prediction.
We plan to explore this in detail in future work.

\begin{figure}[t!]
    \centering
    \includegraphics{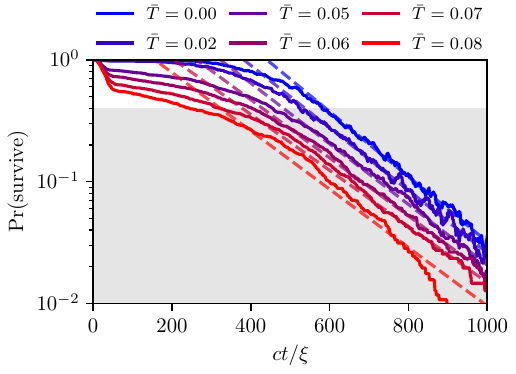}
    \caption{\label{fig:survival-frac-finite-temp}
    Survival probability as a function of dimensionless temperature $\bar{T}=k_\mathrm{B}T/(gn)$, for $\bar{n}=40$.
    For our fiducial parameters this can be translated into a physical temperature using Eq.~\eqref{eq:barT}.
    The decay rates (extracted by fitting in the gray shaded region, shown here as dashed lines) are consistent with being temperature-independent up to roughly $\bar{T}\approx0.06$; beyond this point, large fluctuations in the total modes couple to the relative modes and ruin the effective relativistic picture.}
\end{figure}

\section{Finite-temperature effects}\label{sec:finite-temp}

Thus far we have considered only zero-temperature states of the analog system.
However, any realistic experiment will inevitably be at some finite temperature, and will therefore contain \emph{thermal} as well as quantum fluctuations.
These are potentially a nuisance factor in studying quantum vacuum decay, giving an excess contribution to the decay rate and altering the phenomenology of the nucleated bubbles~\cite{Linde:1981zj}.
It is therefore valuable to estimate the temperature threshold at which these deviations from the zero-temperature case become significant, as this can then guide the development and interpretation of the analog experiments.

\begin{figure}[t!]
    \centering
    \includegraphics{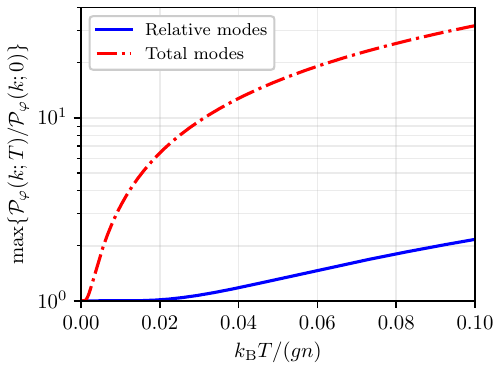}
    \caption{\label{fig:power-finite-temp}
    Enhancement in the fluctuation power spectra of the total and relative phonons as a function of temperature.
    The vertical axis shows the ratio between the finite-temperature and zero-temperature power spectra evaluated at the minimum wavenumber $k_\mathrm{IR}=\uppi/L\approx6.28\times10^{-3}\,\xi^{-1}$, for which the enhancement is maximized.}
\end{figure}

In the framework of the truncated Wigner approximation, we can model the thermal bath by including additional fluctuation power in our initial conditions.\footnote{%
    Other prescriptions and theoretical frameworks exist, including modeling the effects of the thermal bath by adding a stochastic driving term to the Gross-Pitaevskii equations~\cite{Gardiner:2002spg,Billam:2020xna,Billam:2021qwt,Billam:2022ykl}.
    However, our treatment here allows us to model quantum and thermal fluctuations in a simple and conceptually unified way.
    A detailed comparison against alternative simulation methods would be interesting, but is beyond our present scope.}
This amounts to replacing vacuum expectation values with traces over a thermal density matrix, resulting in a scale-dependent enhancement to the relative phase power spectrum,
    \begin{equation}
    \label{eq:power-finite-temp}
        \mathcal{P}_\varphi(k;T)=\coth\qty(\frac{\hbar\omega_k}{2k_\mathrm{B}T})\,\mathcal{P}_\varphi(k;0),
    \end{equation}
    as well as for the relative density, and the total phase and density.
(Here $\coth x=(1+\rme^{-2x})/(1-\rme^{-2x})$ is the hyperbolic cotangent function.)
It is convenient to work in terms of the dimensionless temperature,
    \begin{equation}
    \label{eq:barT}
        \bar{T}=\frac{k_\mathrm{B}}{gn}T\approx\frac{T}{182\,\mathrm{nK}},
    \end{equation}
    where the numerical value corresponds to our particular choice of experimental parameters (cf. Table~\ref{tab:parameters}).

Fig.~\ref{fig:survival-frac-finite-temp} shows the survival probability in ensembles of simulations at various temperatures, with $\bar{n}=40$.
For dimensionless temperatures $\bar{T}\lesssim0.06$ we see that, notwithstanding some differences in the initial nonexponential transient phase, the exponential decay rates are all consistent with the zero-temperature result.
At higher temperatures, rather than finding an enhanced rate of relativistic decays, we instead find that the exponential decay model becomes an increasingly poor fit to the empirical survival probabilities.
We interpret this finding as indicating the breakdown of the relativistic analogy at high temperatures, and conjecture that this breakdown is due to the impact of thermal noise on the \emph{total} phonon modes.
In contrast to the relative modes, which have an effective mass $m_\mathrm{fv}$ due to the potential barrier around the false vacuum, the total modes have a massless dispersion relationship $\omega_k\simeq ck$ in the IR, allowing them to become excited to very large amplitudes by the thermal bath, as illustrated in Fig.~\ref{fig:power-finite-temp}.
The coupling between the total and relative modes then becomes significant, and spoils the effective relativistic dynamics of the relative modes.
As evidence for this interpretation, we note that the $\bar{T}\lesssim0.06$ threshold determined empirically from our simulations is just below the theoretically-predicted threshold at which the total modes of this system should lose phase coherence, $\bar{T}_\phi=\bar{n}/\bar{L}=0.08$~\cite{Billam:2020xna}.

Our results show that dimensionless temperatures of $\bar{T}\lesssim0.06$ should give us access to a setting closely resembling the zero-temperature dynamics of the analog vacuum decay process.
This translates into physical temperatures of $T\lesssim10.9\,\mathrm{nK}$ for our proposed parameters.
Note that our interpretation in terms of the phase coherence temperature $\bar{T}_\phi=\bar{n}/\bar{L}$ implies that this threshold should scale proportionally with the fluctuation-amplitude parameter $\bar{n}$, so that the $T\lesssim10.9\,\mathrm{nK}$ benchmark should be viewed as a \emph{minimal} requirement, with lower temperatures giving us access to vacuum decay rates over a broader range of parameter space.
This benchmark is readily accessible with current experimental setups, which routinely reach temperatures on the order of a few nK, and have even recorded temperatures as low as tens of pK~\cite{Deppner:2021fks}.

\section{Summary and outlook}\label{sec:summary}

Quantum analog experiments present a powerful new tool for understanding relativistic vacuum decay.
Here we have carried out a detailed study of one such proposed experimental setup, which uses a rapidly modulated coupling between two atomic Bose-Einstein condensates to engineer a metastable false vacuum state for the relative phase.
We have derived the spectrum of quantum fluctuations around this state, and have shown that this spectrum asymptotically matches that of the effective Klein-Gordon field in the IR.

As well as providing further evidence for the suitability of the cold-atom analog for studying relativistic physics, this vacuum fluctuation spectrum is also a crucial input for semiclassical lattice simulations of this system.
By carrying out a suite of such simulations, we have confirmed the key theoretical expectations for the analog false vacuum: that it undergoes exponential decay, at a rate that is exponentially sensitive to the amplitude of the vacuum fluctuations.
We have also shown that using an alternative fluctuation spectrum --- in this case, white noise, which has been used in several previous studies of this system --- leads to an enhanced decay rate compared to the pseudorelativistic vacuum fluctuations, as this corresponds to putting the system in an excited initial state.

In carrying out these simulations, we have identified a realistic set of parameters that will allow us to study vacuum decay with current experimental capabilities.
This includes a protocol for scanning over fluctuation amplitudes, and thus decay rates, while keeping all other natural scales of the system fixed, enabling detailed and controlled experimental studies of the decay rate.

As well as the zero-temperature fluctuation spectrum, we have derived the enhancement of the fluctuation power due to thermal noise at finite temperature.
We find that, so long as the system is below a given temperature threshold (which we argue is set by the coupling between the total and relative phase degrees of freedom), the decay rate extracted from our simulations is consistent with that at zero temperature.
For our proposed parameters, this threshold lies well within reach of current experiments, meaning that we should be able to empirically test the physics of \emph{quantum} bubble nucleation in the near future.

Our results here rely on several simplifying assumptions, which we plan to relax in future work.
In particular, we have treated the BEC system as periodic, neglecting boundary effects due to the external trapping potential.
In a forthcoming companion paper, we will generalize our Bogoliubov analysis to derive the \emph{inhomogeneous} vacuum fluctuations in a box trap, and investigate the impact of these boundary effects on the bubble nucleation rate.
We have also neglected in our calculations the backreaction of the fluctuations onto the mean-field dynamics of the BEC, and corresponding renormalization of the bare parameters of the effective relativistic theory.
Incorporating these effects should allow for a more precise understanding of the validity of the relativistic analogy, improve the initialization and interpretation of our lattice simulations, and enable more detailed comparisons with instanton predictions.
These developments will enable the first experimental tests of relativistic vacuum decay.

\begin{figure*}[t!]
    \centering
    \includegraphics{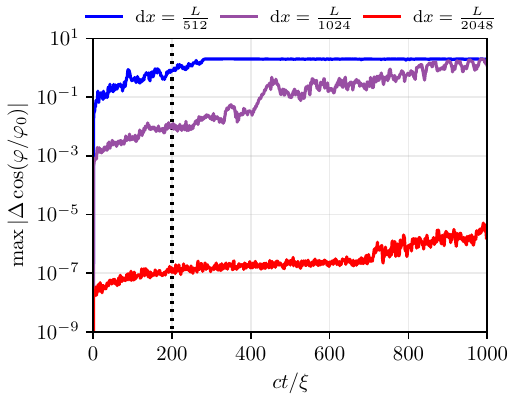}
    \includegraphics{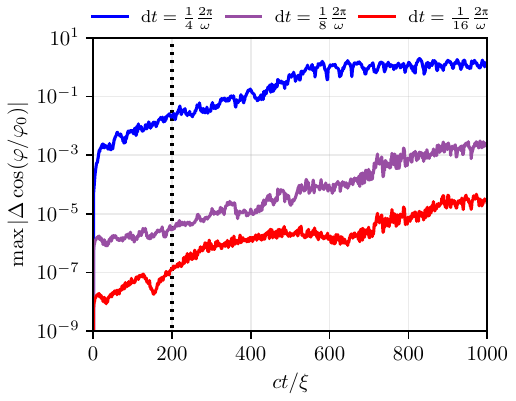}
    \caption{\label{fig:convergence}
    Pointwise convergence of our numerical solutions for increasing spatial and temporal resolution (left and right panels, respectively) in simulations with $\bar{n}=30$.
    Each curve shows the maximum absolute pointwise difference in $\cos(\varphi/\varphi_0)$ between one solution with the stated resolution and another with double the spatial or temporal resolution, starting from identical initial conditions.
    The vertical dotted lines show the time of bubble nucleation in the converged simulations.
    Note that the resolution used in our simulations discussed in Secs.~\ref{sec:lattice} and~\ref{sec:finite-temp} corresponds to the red curves here.}
\end{figure*}

\begin{acknowledgments}
    We thank Tom Billam, Nishant Dogra, Christoph Eigen, Zoran Hadzibabic, Konstantinos Konstantinou, Ian Moss, and Dalila Pîrvu for valuable discussions.
    This work was supported by the Science and Technology Facilities Council through the UKRI Quantum Technologies for Fundamental Physics Programme [grant number ST/T005904/1], and was partly enabled by the UCL Cosmoparticle Initiative.
    This work used computing equipment funded by the Research Capital Investment Fund (RCIF) provided by UKRI, and partially funded by the UCL Cosmoparticle Initiative.
    This work used facilities provided by the UCL Cosmoparticle Initiative.
    The work of JB was partially supported by the Natural Sciences and Engineering Research Council of Canada (NSERC).
    JB was supported in part by the Simons Modern Inflationary Cosmology program.
    MCJ is supported by the National Science and Engineering Research Council through a Discovery grant.
    SW acknowledges support provided by the Leverhulme Research Leadership Award (RL2019- 020), the Royal Society University Research Fellowship (UF120112, RF{\textbackslash}ERE{\textbackslash}210198) and the Royal Society Enhancements Awards and Grants (RGF{\textbackslash}EA{\textbackslash}180286, RGF{\textbackslash}EA{\textbackslash}181015), and partial support by the Science and Technology Facilities Council (Theory Consolidated Grant ST/P000703/1).
    This research was supported in part by Perimeter Institute for Theoretical Physics.
    Research at Perimeter Institute is supported in part by the Government of Canada through the Department of Innovation, Science and Economic Development Canada and by the Province of Ontario through the Ministry of Colleges and Universities.
    We acknowledge the use of the Python packages NumPy~\cite{Harris:2020xlr}, SciPy~\cite{Virtanen:2019joe}, and Matplotlib~\cite{Hunter:2007ouj}.
    The data that support the findings of this study are available from the corresponding author, ACJ, under reasonable request.
\end{acknowledgments}

\section*{Author contributions}

Contributions based on the CRediT (Contribution Roles Taxonomy) system.

\textbf{ACJ}: conceptualization; methodology; software; formal analysis; investigation; data curation; interpretation and validation;  visualization; writing (original draft).
\textbf{JB}: conceptualization; methodology; software; formal analysis; interpretation and validation; writing (review and editing).
\textbf{HVP}: conceptualization; methodology; interpretation and validation; writing (review and editing); funding acquisition.
\textbf{AP}: conceptualization; methodology; interpretation and validation; writing (review and editing); funding acquisition.
\textbf{MCJ}: conceptualization; methodology; interpretation and validation; writing (review).
\textbf{SW}: conceptualization; interpretation and validation; writing (review); funding acquisition.

\appendix
\section{Bogoliubov analysis}\label{sec:bogoliubov}

The Hamiltonian~\eqref{eq:quadratic-hamiltonian} is diagonalized by applying a Bogoliubov transformation to the atomic field modes,
    \begin{equation}
        \hat{a}^{\phantom{\dagger}}_{\vb*k}=-\rmi(u_k\hat{\psi}^-_{\vb*k}+v_k\hat{\psi}^{-\dagger}_{\vb*k}),
    \end{equation}
    where the coefficients are given by
    \begin{align}
    \begin{split}
    \label{eq:bogoliubov-coefficients}
        u_k^2&=\frac{1}{2}\qty[\frac{gn}{2\hbar\omega_k}(\xi^2k^2+2-4\epsilon)+1],\\
        v_k^2&=\frac{1}{2}\qty[\frac{gn}{2\hbar\omega_k}(\xi^2k^2+2-4\epsilon)-1],
    \end{split}
    \end{align}
    with $\omega_k$ the dispersion relation given in Eq.~\eqref{eq:dispersion-relation}.
Since the condition $u_k^2-v_k^2=1$ is satisfied, one can verify that these normal modes obey the standard bosonic commutation relations,
    \begin{equation}
        \comm*{\hat{a}^{\phantom{\dagger}}_{\vb*k}}{\hat{a}^\dagger_{\vb*k'}}=\delta_{\vb*k,\vb*k'},\qquad\comm*{\hat{a}^{\phantom{\dagger}}_{\vb*k}}{\hat{a}^{\phantom{\dagger}}_{\vb*k'}}=\comm*{\hat{a}^\dagger_{\vb*k}}{\hat{a}^\dagger_{\vb*k'}}=0.
    \end{equation}
This, combined with the diagonalized Hamiltonian~\eqref{eq:diagonalized-hamiltonian}, allows us to interpret $\hat{a}^{\phantom{\dagger}}_{\vb*k}$ and $\hat{a}^\dagger_{\vb*k}$ as ladder operators for a set of independent harmonic oscillators, one for each normal mode.
The false vacuum $\ket{\Omega_\mathrm{fv}}$ is then naturally defined as the ground state of these oscillators.

Inserting the normal modes into Eq.~\eqref{eq:varphi-relative-modes}, we find that the Fourier modes of the relative phase can be written as
    \begin{equation}
        \hat{\varphi}_{\vb*k}^{\phantom{\dagger}}=\frac{\hbar c}{2\sqrt{gn}}(u_k+v_k)(\hat{a}_{\vb*k}+\hat{a}_{-\vb*k}^\dagger).
    \end{equation}
In the IR ($\xi k\ll1$), this corresponds exactly to the equivalent expression for a canonically-normalized Klein-Gordon field~\cite{Peskin:1995ev},
    \begin{equation}
        \hat{\varphi}_{\vb*k}^{\phantom{\dagger}}\simeq\sqrt{\frac{\hbar c^2}{2\omega_k}}(\hat{a}_{\vb*k}+\hat{a}_{-\vb*k}^\dagger),
    \end{equation}
    which automatically guarantees that all expectation values will match those of the Klein-Gordon case in this regime, including the power spectrum~\eqref{eq:kg-spectrum}.
To simulate white-noise initial conditions, we simply replace the coefficients in Eq.~\eqref{eq:bogoliubov-coefficients} with $u_k=1$ and $v_k=0$.

In our lattice simulations, we represent the normal modes $\hat{a}_{\vb*k}$ as classical stochastic variables $a_{\vb*k}$, with expectation values defined by symmetrizing over classically-equivalent operator orderings; e.g.,
    \begin{equation}
        \ev*{\abs{a_{\vb*k}}^2}=\frac{1}{2}\ev{\hat{a}^{\phantom{\dagger}}_{\vb*k}\hat{a}^\dagger_{\vb*k}+\hat{a}^\dagger_{\vb*k}\hat{a}^{\phantom{\dagger}}_{\vb*k}}{\Omega_\mathrm{fv}}=\frac{1}{2}.
    \end{equation}
A simple calculation then shows that each $a_{\vb*k}$ is an independent draw from a circularly-symmetric complex Gaussian distribution, with real and imaginary parts each having variance $1/4$.
In the finite-temperature case, this variance is enhanced by a factor of $\coth(\hbar\omega_k/2k_\mathrm{B}T)$.

Notice that, while $a_{\vb*k}$ and $a_{-\vb*k}$ are statistically independent, the Bogoliubov transformation mixes positive and negative momenta together so that $\psi_{\vb*k}$ and $\psi_{-\vb*k}$ are \emph{not} independent.
Initializing $\psi_{\vb*k}$ and $\psi_{-\vb*k}$ independently leads to nontrivial correlations between $a_{\vb*k}$ and $a_{-\vb*k}$, and therefore fails to correctly capture the statistics of the false vacuum state.

\begin{figure}[t!]
    \centering
    \includegraphics{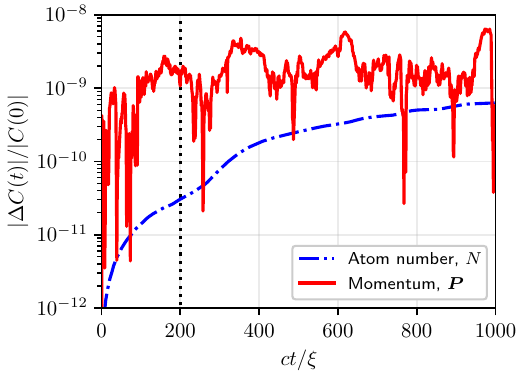}
    \caption{\label{fig:noether-charges}
    Relative variations in the Noether charges~\eqref{eq:noether-charges} for the same simulation shown in the red curves of Fig.~\ref{fig:convergence}.}
\end{figure}

\section{Numerical methods and convergence tests}\label{sec:numerics}

Our code solves the classical equations of motion for the atomic fields $\vb*\psi(\vb*x,t)=(\psi_1,\psi_2)^\mathsf{T}$, corresponding to the time-modulated Hamiltonian~\eqref{eq:hamiltonian-time-dependent},
    \begin{equation}
        \rmi\hbar\partial_t\vb*\psi=\mathcal{O}\vb*\psi,
    \end{equation}
    where we define the differential operator,
    \begin{align}
    \begin{split}
        \mathcal{O}(\vb*x,t)&=\mathcal{O}_\mathrm{lin}+\mathcal{O}_\mathrm{nlin}\\
        \mathcal{O}_\mathrm{lin}(\vb*x,t)&=-\frac{\hbar^2}{2m}\laplacian-\hbar\nu(t)\pmqty{0 & 1 \\ 1 & 0}-\mu_\mathrm{lin},\\
        \mathcal{O}_\mathrm{nlin}(\vb*x,t)&=g\pmqty{\abs{\psi_1(\vb*x,t)}^2 & 0 \\ 0 & \abs{\psi_2(\vb*x,t)}^2}-\mu_\mathrm{nlin},
    \end{split}
    \end{align}
    which we have split into a linear and a nonlinear piece.
Each piece has its own chemical potential, which can be chosen for convenience --- e.g., to minimize sinusoidal oscillations in the homogeneous mode of the total phase --- as these have no effect on the relative phase $\varphi$.
Evolution under each of these operators individually can be solved \emph{exactly}; the nonlinear piece conserves the amplitude of each field and simply performs a local phase rotation,
    \begin{equation}
        \vb*\psi(\vb*x,t)=\exp[-\rmi\frac{t-t_0}{\hbar}\mathcal{O}_\mathrm{nlin}(\vb*x,t_0)]\vb*\psi(\vb*x,t_0),
    \end{equation}
    while the linear piece can be solved by going to Fourier space,
    \begin{align}
    \begin{split}
    \label{eq:linear-exact-sol}
        \vb*\psi(\vb*x,t)&=\mathcal{F}^{-1}_{\vb*k\to\vb*x}\Bigg\{\exp[-\rmi\frac{t-t_0}{\hbar}\qty(\frac{\hbar^2k^2}{2m}-\mu_\mathrm{lin})]\\
        &\times\pmqty{\cos R(t,t_0) & \rmi\sin R(t,t_0) \\ \rmi\sin R(t,t_0) & \cos R(t,t_0)}\mathcal{F}_{\vb*x\to\vb*k}\qty{\vb*\psi(\vb*x,t_0)}\Bigg\},\\
        R(t,t_0)&=\epsilon gn\frac{t-t_0}{\hbar}+\lambda\sqrt{\epsilon/2}[\sin(\omega t)-\sin(\omega t_0)],
    \end{split}
    \end{align}
    where $\mathcal{F}_{\vb*x\to\vb*k}$ represents a Fourier transform, and $\mathcal{F}^{-1}_{\vb*k\to\vb*x}$ its inverse.
(These are implemented numerically as fast Fourier transforms, so that in practice Eq.~\eqref{eq:linear-exact-sol} is only exact under the assumption that the fields are band-limited with maximum wavenumber less than or equal to the Nyquist frequency on the lattice.)

While there is no exact solution for the evolution under $\mathcal{O}=\mathcal{O}_\mathrm{lin}+\mathcal{O}_\mathrm{nlin}$ from generic initial data, we can approximate this full evolution by chaining together a series of short steps with each of the individual operators,
    \begin{align}
    \begin{split}
        \vb*\psi(\vb*x,&t_0+\updelta t)=\rme^{-\rmi a_1\mathcal{O}_\mathrm{lin}\frac{\updelta t}{\hbar}}\rme^{-\rmi b_1\mathcal{O}_\mathrm{nlin}\frac{\updelta t}{\hbar}}\times\cdots\\
        &\times\rme^{-\rmi a_k\mathcal{O}_\mathrm{lin}\frac{\updelta t}{\hbar}}\rme^{-\rmi b_k\mathcal{O}_\mathrm{nlin}\frac{\updelta t}{\hbar}}\vb*\psi(\vb*x,t_0)+\order{\updelta t^{n+1}},
    \end{split}
    \end{align}
    where the dimensionless coefficients $a_i$, $b_i$, ($i=1,\ldots,k$) are chosen such that the integrator is exact to order $n$ in the small timestep $\updelta t$.
Integrators of this form are symplectic, in the sense that they exactly conserve phase space volume.
We implement an efficient realization of this integrator from \citet{Yoshida:1990zz}, which uses $k=16$ steps and is accurate to order $n=8$.

In Fig.~\ref{fig:convergence} we show convergence tests of our code for increasing spatial and temporal resolution, measuring numerical errors in terms of pointwise differences in the cosine of the relative phase field, $\cos(\varphi/\varphi_0)$.
For the level of resolution used in our simulations in Secs.~\ref{sec:lattice} and~\ref{sec:finite-temp}, we see that the maximum error is on the order of $\sim10^{-7}$ prior to bubble nucleation, and at most $\sim10^{-5}$ even long after bubble nucleation.
This indicates that our simulations are numerically converged, even in the highly dynamical nonlinear regime.

We also test our code by checking for violations in conservation of the Noether charges associated with the cold-atom Hamiltonian~\eqref{eq:hamiltonian-time-dependent},
    \begin{align}
    \begin{split}
    \label{eq:noether-charges}
        N&=\int_V\dd{\vb*x}\sum_i\abs{\psi_i}^2,\\
        \vb*P&=\int_V\dd{\vb*x}\sum_i\frac{\rmi}{2}\qty(\psi_i\grad\psi_i^*-\psi_i^*\grad\psi_i),
    \end{split}
    \end{align}
    which correspond to the total number of atoms and the total momentum of the system, respectively~\cite{Braden:2019vsw}.
(Note that the total energy is \emph{not} exactly conserved, due to the explicit time-dependence of the rf modulation term in the Hamiltonian.)
As shown in Fig.~\ref{fig:noether-charges}, both charges are conserved to the level of a few parts per billion in simulations at our fiducial resolution.

\bibliography{analog-fvd}
\end{document}